\documentclass{emulateapj}
\usepackage{color,graphicx}

\begin{document}

\title{The Hot and Cold Spots in the \emph{Wilkinson Microwave Anisotropy Probe} Data are Not Hot and Cold Enough}
\author{David L. Larson} 
\email{dlarson1@uiuc.edu}
\affil{Department of Physics, University of Illinois at Urbana-Champaign, Urbana, IL 61801, USA}
\author{Benjamin D. Wandelt\footnote{2003/2004 NCSA faculty fellow.}}
\email{bwandelt@uiuc.edu}
\affil{Department of Physics, University of Illinois at Urbana-Champaign, Urbana, IL 61801, USA \\
Department of Astronomy, University of Illinois at Urbana-Champaign, Urbana, IL 61801, USA}

\begin{abstract}

This Letter presents a frequentist analysis of the hot and cold spots of the
cosmic microwave background data collected by the \emph{Wilkinson Microwave
Anisotropy Probe} (WMAP).  We compare the WMAP temperature statistics of extrema
(number of extrema, mean excursion, variance, skewness and kurtosis of the
excursion) to Monte Carlo simulations.  We find that on average, the local
maxima (high temperatures in the anisotropy) are too cold and the local minima
are too warm.  
In order to quantify this claim we describe a two-sided statistical 
hypothesis test which we advocate for other investigations of the Gaussianity 
hypothesis. Using this test we reject the isotropic Gaussian hypothesis at 
more than 99\% confidence in a well-defined way. Our claims are based only on 
regions that are outside the most conservative WMAP foreground mask. 
We perform our test separately on maxima and minima, and on the north and 
south ecliptic and Galactic hemispheres and reject Gaussianity at above 95\% 
confidence for almost all tests of the mean excursions. The same test also 
shows the variance of the maxima and minima to be low in the ecliptic north 
(99\% confidence) but consistent in the south; this effect is not as 
pronounced in the Galactic north and south hemispheres.

\end{abstract}
\keywords{cosmic microwave background}

\section{Introduction}

The \emph{Wilkinson Microwave Anisotropy Probe} 
(WMAP) data provide the most detailed data on the full sky cosmic microwave background (CMB) to date.
This information about the initial density fluctuations in the universe
allows us to test the cosmological standard model at unprecedented levels of detail.
(Bennett {et~al.} 2003a)
A question of fundamental importance to our understanding 
of the origins of these primordial seed perturbations 
is whether the CMB radiation 
is really an isotropic and Gaussian random field, as generic inflationary theories 
predict (Starobinsky 1982; Guth \& Pi 1982; Bardeen {et~al.} 1983).

A natural way to study the CMB is to look at the local extrema.  This
was initially suggested because the high signal-to-noise ratio
at the hot spots means they would be detected 
first 
(Sazhin 1985; Zabotin \& Naselsky 1985; Vittorio \& Juszkiewicz 1987; Bond \& Efstathiou 1987).  
Heavens \& Sheth calculate analytically the two-point correlation
function of the local extrema  (Heavens \& Sheth 1999).
In addition, extrema trace the topological properties of the temperature map; this makes 
them good candidates for study (Wandelt {et~al.} 1998).

We pursue this investigation by simulating Gaussian Monte Carlo CMB skies and comparing the WMAP data
to those simulations.
We choose several statistics and then check to
see if the WMAP statistics lie in the middle of the Monte Carlo distributions of statistics.
We present results on the one-point functions of the local extrema: their number,
mean excursion, and variance, and skewness and kurtosis of the excursion.  

The literature contains many other searches for non-Gaussianity, in the WMAP data and other CMB experiments.
For example, Vielva et~al.\ detect non-Gaussianity in the three- and four-point wavelet moments 
 (Vielva et al.\ 2004), Chiang et~al.\ detect it in phase correlations 
between spherical harmonic coefficients (Chiang et al.\ 2003; see also 
Chiang et al.\ 2002, 2004),
and Park finds it in the genus Minkowski functional  (Park 2004).
Eriksen et~al.\ find anisotropy in the $n$-point
functions of the CMB in different patches of the sky  (Eriksen et al.\ 2004). 
Others discuss possible methods of detecting non-Gaussianity. 
Aliaga et~al.\ look at studying non-Gaussianity through spherical wavelets and ``smooth tests of 
goodness-of-fit''  (Aliaga et al.\ 2003).
Cabella et~al.\ review three methods of studying non-Gaussianity: through Minkowski functionals,
spherical wavelets, and the spherical harmonics  (Cabella et al.\ 2004). They propose a way to combine
these methods.

Komatsu et~al.\ discuss a fast way to test the bispectrum for primordial non-Gaussianity in the CMB  (Komatsu 2003a), 
and do not detect it  (Komatsu et al.\ 2003b).
Finally, Gazta{\~n}aga et al.\ find the CMB to be consistent with Gaussianity when considering the
two and three-point functions  (Gazta{\~n}aga \& Wagg 2003; Gazta{\~n}aga et al.\ 2003).
To this work, we add a strong detection of non-Gaussianity based on generic features: the local extrema.

The Letter is laid out as follows.  
The next section discusses our method for making Monte Carlo simulations of the CMB sky and calculating
statistics on both the simulations and the WMAP data.  It also explains our statistical tests.
Section 3 describes our results.
We conclude in section 4.

\section{Method}

We test the WMAP data of the CMB sky by comparing the one-point statistics of its extrema to those same statistics on
several sets of Monte Carlo--simulated Gaussian skies.   
Our null hypothesis is that the statistics of the WMAP data are drawn from the same probability density function (PDF)
as the statistics of the Monte Carlo skies.
If some WMAP one-point statistic 
falls lower or higher than most of the Monte Carlo statistics, this indicates 
that our hypothesis may be false.

We examine several inputs to our Monte Carlo simulation to see how those change
the Monte Carlo distribution of one-point statistics around the WMAP one-point statistics.
We start very generally, looking at 
different frequency bands and Galactic masks,
and then narrow our search.
Initially, we look at simulations including the three frequency bands (Q, V, and W) 
and the three published Galactic masks for the WMAP data.
Then we check to see if changing to a different published theoretical power spectrum 
affects our results.
Finally, we look for anisotropy between the statistics of the ecliptic and Galactic north and south
hemispheres. 

\subsection{Monte Carlo Simulation}

A general outline of our Monte Carlo simulation process follows.
Each set of skies is labeled by its theoretical 
power spectrum,
frequency band (Q, V, or W), and Galactic mask.
The frequency band determines both the 
(azimuthally averaged) beam shape function and the noise properties on the sky.
The simulated CMB skies are created as follows:
\begin{enumerate}
\item A Gaussian CMB sky is created with SYNFAST, 
	using a power spectrum and a beam function.
	The HEALPix\footnote{See http://www.eso.org/science/healpix/} 
	pixelization of the sphere is used, with $N_{side}$ = 512.
\item Random Gaussian noise is added to the sky (at each pixel) 
	according to the published noise characteristics
	of the band being simulated.  The WMAP radiometers are characterized as having white Gaussian 
	noise (Jarosik {et~al.} 2003).
\item The monopole and dipole moments of the sky (outside of the chosen Galactic mask) are removed.
\end{enumerate}
We make no attempt to simulate any foregrounds, including the galaxy; our analysis ignores
data inside a Galactic mask and uses the cleaned maps published on LAMBDA 
(Legacy Archive for Microwave Background Data Analysis; NASA 2003).

For each Monte Carlo set, one of four power spectra is used.  These
are the power spectra published by the WMAP team on LAMBDA (NASA 2003).
We primarily use the best-fit (bf)
theoretical power spectrum to a cold dark matter universe with a running spectral index
using the WMAP, Cosmic Background Imager (CBI), Arcminute Cosmology Bolometer Array Receiver (ACBAR), 
Two-Degree Field, and Ly$\alpha$ data.  
In addition, we check the unbinned power spectrum (w) directly measured by WMAP, 
the power law (pl) theoretical power spectrum fit to WMAP, CBI and ACBAR, 
and a running index (ri) theoretical power spectrum fit to WMAP, CBI and ACBAR. 
See Spergel {et~al.} (2003), Bennett {et~al.} (2003a), and  NASA (2003)
for more information.

The Galactic masks used are the Kp0, Kp2, and Kp12 masks
published by the WMAP team (Bennett {et~al.} 2003b).  
To check for differences between the north and south ecliptic hemispheres, we define additional
masks that extend the Kp0 Galactic mask to block either the north or south hemisphere as well.
For example, the ecliptic south (ES) mask blocks the northern ecliptic sky as well as the galaxy.
As a control, we also extend the Kp0 mask for Galactic north and south hemispheres (GN and GS) to bring the 
total number of masks up to seven: Kp0, Kp2, Kp12, GS, GN, ES, EN.
We use the same masking and dipole removal procedure for the WMAP data as for the Monte Carlo skies.

The WMAP data that we use are the cleaned, published maps.  They are published by channel, 
so we calculate an \emph{unweighted} average over (for example) all four W-band channels to 
get a map for the W band.  The noise variance is calculated accordingly.
We compute an unweighted average of the maps so that
we can combine the WMAP beam functions through a simple average.  

\subsection{Analysis and Hypothesis Test}

Our analysis of both the Monte Carlo and WMAP skies involves the following:
We find the local maxima and minima of the HEALPix grid using HOTSPOT.
Then we discard the extrema blocked by the Galactic mask.
We calculate the statistics (number, mean, variance, skewness, and kurtosis) 
on the temperatures of the maxima and minima,
and then statistically analyze the significance of the position of the WMAP statistic
among the Monte Carlo statistics.
Because we consider only the one-point statistics,
we consider only the temperature values, not their locations.

We calculate our five statistics for the maxima and minima separately.
The two statistics which are typically negative for the minima, the mean and skewness, 
are multiplied by $-1$ in our results,
to make comparison with the maxima statistics more clear. 

For the rest of this section, we explain our analysis of the statistics in detail.
To simplify the discussion, we consider the analysis of only one statistic on either maxima or minima,
as we analyze the results for each statistic separately.

Our Monte Carlo simulations are binomial trials, where the statistic calculated on 
a simulation can lie either above or below the WMAP statistic.
It lies below the WMAP statistic with probability $p$, and for some set of $n$ trials,
$i$ of the trials will have statistics below the WMAP statistic.
Given $p$, the probability of $i$ is 
$P(i|p)=[n!/i!(n-i)!]p^i(1-p)^{n-i}$.
The value $\hat{p}\equiv i/n$ is both an unbiased and maximum likelihood
estimator of $p$.

We are interested in whether $p$ is near 0 or 1, 
since that indicates that our hypothesis---that the WMAP statistic 
came from the same PDF as the Monte Carlo statistics---may be false.
Because we do not have an alternative distribution for the WMAP statistic
that we can test against the Monte Carlo distribution, 
we do not test our hypothesis as phrased.  
We only look at a hypothesis $H_0$ that claims $p$ is in some interval, 
$p\in (\alpha/2, 1-\alpha/2)$, where we have arbitrarily chosen $\alpha = 0.05$.

We devise a statistical test of this hypothesis.  
Given our experimental result $i$, 
we construct the $1-\alpha=95\%$ symmetric confidence interval for $p$, as described in
Kendall \& Stuart (1973).  
If this confidence interval lies
entirely within the interval $[0, \alpha/2]$ or entirely within $[1-\alpha/2, 1]$ then we 
reject our hypothesis, $H_0$.  
We reject $H_0$ for no values of $i$ when $n=99$, 
for $0\le i\le 15$ or $985\le i\le 1000$ when $n=1000$, and for 
$0\le i\le103$ or $4897\le i\le 5000$ when $n=5000$.

This interval is a ``95\%'' confidence interval in the following frequentist (non-Bayesian) sense.  
Suppose we repeat the experiment (with the same number of Monte Carlo runs, and the same WMAP data) 
many times and get many values of $i$.  We recalculate the confidence intervals each time, for
each particular value of $i$.
Ninety-five percent of the confidence intervals we calculate will contain the true value of $p$.

Our test is biased in favor of $H_0$.  
Let $H_1$ be the alternative hypothesis $p \in [0,\alpha/2] \cup [1-\alpha/2,1]$.  
Then, for some values of $p$ where $H_1$ is true (for example, $p=0.02$, $n=1000$),
our test will choose $H_0$ more often than $H_1$, given that $i$ is a random variable with
probability $P(i|p)$.
If desired, we can make the test unbiased by 
changing our value of $\alpha$ in the hypotheses $H_0$ and $H_1$, but
keeping the test (range of $i$ for which $H_0$ is accepted) the same. 

For $n=1000$, we have an unbiased test if 
$\alpha = 0.0313$, and  
for $n=5000$, we have an unbiased test if 
$\alpha = 0.0415$.  
Note that these values are less than $\alpha=0.05$.
For any value of $p$, these tests are at least as likely to choose
the correct hypothesis as the incorrect one.
This is a 50\% confidence, as opposed to our previous 95\% confidence.
This interpretation of the test does not change our results; 
it merely provides the different perspective that our test may be considered 
an unbiased 96.9\% test, for $n=1000$;
or an unbiased 95.9\% test, for $n=5000$.

\section{Results}

\begin{figure}
\includegraphics[width=.48\textwidth, keepaspectratio]{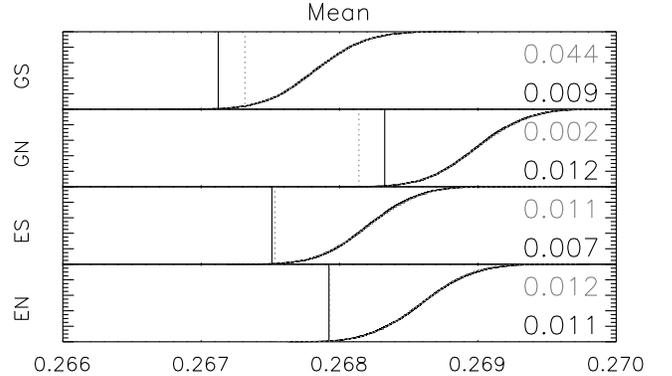}
\caption{Cumulative distribution functions (CDFs) of mean temperature value (in units of millikelvins)
	of the local extrema, found in
	sets of 5000 Monte Carlo simulations for four Galactic masks: GS, GN, ES, and EN.  Best fit power spectrum and
	W-band data are used.
	Means of the minima are negated for comparison.  
	Maxima CDF is dotted while minima CDF is solid.  Note their visual similarity.
	Statistics measured on WMAP data are shown as two vertical lines, dotted for
	maxima and solid for minima.  Numbers on right are the same probabilities as in 
	Table~\ref{bigTable}; for each pair, probability for the maxima is higher on the page. 
	}
\label{mean5000}
\end{figure}

\begin{deluxetable}{llllllr}
\tablecaption{Estimated Probabilities $\hat{p}$ of the CMB Statistics Being Less than
the Value Measured by WMAP, Based on Several Monte Carlo Samplings of Those Statistics
\label{bigTable}
}
\startdata
\hline
\hline
Identifier & 0 & 1 & 2 & 3 & 4 & 5 \\
\hline
bf, Q, Kp0, max & 0.374 & 0.000 & 0.030 & 0.566 & 0.889 &   99\\
bf, Q, Kp0, min & 0.010 & 0.000 & 0.232 & 0.919 & 0.707 & \\
bf, Q, Kp2, max & 0.333 & 0.000 & 0.152 & 0.545 & 0.899 &   99\\
bf, Q, Kp2, min & 0.010 & 0.000 & 0.293 & 0.919 & 0.768 & \\
bf, Q, Kp12, max & 0.020 & 0.000 & 0.576 & 0.929 & 1.000 &   99\\
bf, Q, Kp12, min & 0.000 & 0.000 & 0.798 & 0.919 & 1.000 & \\
bf, V, Kp0, max & 0.091 & 0.616 & 0.182 & 0.495 & 0.848 &   99\\
bf, V, Kp0, min & 0.030 & 0.182 & 0.303 & 0.990 & 0.960 & \\
bf, V, Kp2, max & 0.061 & 0.384 & 0.061 & 0.364 & 0.727 &   99\\
bf, V, Kp2, min & 0.040 & 0.202 & 0.263 & 1.000 & 0.960 & \\
bf, V, Kp12, max & 0.000 & 0.384 & 0.212 & 0.343 & 0.889 &   99\\
bf, V, Kp12, min & 0.020 & 0.232 & 0.444 & 0.980 & 1.000 & \\
bf, W, Kp0, max & 0.475 & 0.000 & 0.141 & 0.364 & 0.475 &   99\\
bf, W, Kp0, min & 0.414 & 0.000 & 0.343 & 0.879 & 0.646 & \\
bf, W, Kp2, max & 0.495 & 0.000 & 0.131 & 0.182 & 0.283 &   99\\
bf, W, Kp2, min & 0.293 & 0.000 & 0.323 & 0.808 & 0.616 & \\
bf, W, Kp12, max & 0.434 & 0.000 & 0.242 & 0.313 & 0.253 &   99\\
bf, W, Kp12, min & 0.364 & 0.010 & 0.505 & 0.707 & 0.737 & \\
\hline
pl, W, Kp0, max & 0.427 & {0.002}* & 0.129 & 0.461 & 0.410 & 1000\\
pl, W, Kp0, min & 0.377 & {0.000}* & 0.224 & 0.880 & 0.704 & \\
ri, W, Kp0, max & 0.388 & {0.000}* & 0.216 & 0.285 & 0.310 & 1000\\
ri, W, Kp0, min & 0.396 & {0.000}* & 0.406 & 0.821 & 0.618 & \\
\hline
w, W, GS, max & 0.633 & 0.023 & 0.362 & 0.045 & 0.304 & 1000\\
w, W, GS, min & 0.685 & {0.007}* & 0.981 & 0.247 & 0.213 & \\
w, W, GN, max & 0.436 & {0.000}* & 0.168 & 0.504 & 0.159 & 1000\\
w, W, GN, min & 0.243 & {0.006}* & 0.060 & 0.832 & 0.610 & \\
w, W, ES, max & 0.607 & {0.010}* & 0.869 & 0.103 & 0.429 & 1000\\
w, W, ES, min & 0.176 & {0.003}* & 0.923 & 0.244 & 0.152 & \\
w, W, EN, max & 0.470 & {0.011}* & 0.019 & 0.416 & 0.119 & 1000\\
w, W, EN, min & 0.702 & {0.005}* & 0.067 & 0.958 & 0.861 & \\
\hline
bf, W, GS, max & 0.603 & 0.044 & 0.240 & 0.152 & 0.434 & 5000\\
bf, W, GS, min & 0.641 & {0.009}* & 0.852 & 0.405 & 0.284 & \\
bf, W, GN, max & 0.371 & {0.002}* & 0.091 & 0.648 & 0.284 & 5000\\
bf, W, GN, min & 0.209 & {0.012}* & 0.035 & 0.883 & 0.697 & \\
bf, W, ES, max & 0.560 & {0.011}* & 0.472 & 0.198 & 0.487 & 5000\\
bf, W, ES, min & 0.151 & {0.007}* & 0.586 & 0.376 & 0.253 & \\
bf, W, EN, max & 0.436 & {0.012}* & {0.002}* & 0.529 & 0.188 & 5000\\
bf, W, EN, min & 0.668 & {0.011}* & {0.011}* & 0.960 & 0.869 & \\
\hline
\multicolumn{7}{c}{
\begin{minipage}[t]{3.4in}
	Notes.---The identifier column provides the power spectrum, band, and mask used and whether the
	statistics are for minima or maxima.  Power spectra are best fit (bf), power law (pl), running
	index (ri), or measured unbinned WMAP (w).
	Columns labeled 0 through 4 give an unbiased estimate of
	the WMAP statistic's position among the sorted Monte Carlo sample statistics. The statistic
	in column 0 is number of hot spots. The other columns correspond to: 1, mean; 
	2, variance; 3, skewness; and 4, kurtosis
	of extrema temperature values.  (For minima, mean and skewness statistics are 
	negated before estimating probability of WMAP statistic being lower.)
	Column 5 gives the number of Monte Carlo samples
	calculated.  
	Probabilities that indicate that, ``the true value of $p$ is at least 95\% likely 
	to be within 0.025 of either 0 or 1,'' are marked with an asterisk.
	The table shows that the data fall low in the mean temperature distribution 
	for almost every set of simulations.
\end{minipage}
}
\end{deluxetable}

We display our results in Figure~\ref{mean5000} and Table~\ref{bigTable}.  
The figure shows where the means of the WMAP maxima (and minima) lie in the Monte Carlo cumulative distribution functions 
for that statistic.
The table contains our
estimates $\hat{p}$ of $p$.  When the result rejects the hypothesis $H_0$, 
this is noted with a footnote.

We find that the mean temperature of the WMAP extrema, and in some cases the variance, differs significantly from
the simulations, but number of extrema, skewness, and kurtosis are modeled fairly 
well by the simulations.  There is ecliptic north-south asymmetry in
the variance of the extrema.  

For the mean, all of our results reject our hypothesis $H_0$ for the power law and
running index power spectra, and 
all four of our results for the ecliptic north and south hemispheres reject $H_0$
when $n=1000$ and $n=5000$.  Since the statistics for the minima are negated, this means
that the WMAP maxima are too cold and the WMAP minima are too hot.  
This is a very significant result, regardless of whether our tests are considered 95\% tests
biased away from detection, or 
a 96.9\%, and a 95.9\% test.

Even more significant are two 99\% ($\alpha = 0.01$) level detections. In Table~\ref{bigTable}, they are
rows bf, W, GN, max, and mean, and bf, W, EN, max, and variance.  For this level of detection, we only accept
values of i where $0\le i\le 12$ or $4988\le i\le 5000$.  
No 99\% confidence detection was possible with only 1000 iterations.

Using our initial 99 simulations, we find qualitatively low mean excursions in the Q and W bands, but not
the V band.
We chose the W band to examine further because it had the best signal to noise ratio,
and it had the least chance of foregrounds outside the Kp0 mask that we use in our final analysis.

It has been suggested (Eriksen {et~al.} 2004) that there is statistical anisotropy between the ecliptic north
and south hemispheres.  
We see this in the variance of the extrema temperatures.
The ecliptic north hemisphere
has a statistically low variance (in one case at the 99\% level) while the ecliptic south is normal.  
To compare, we find the Galactic north to be slightly low while 
the south is again normal.

\section{Conclusion}

In this Letter we generate simulated CMB skies.  We choose several statistics, and calculate
them on both the simulations and the WMAP sky. 
We hypothesize that the WMAP statistics are drawn from the same distribution
as the simulations' statistics, since we have attempted to accurately simulate the CMB 
sky.
If the WMAP statistic is higher or lower than most of the simulations' statistics, 
this indicates that the WMAP statistic's underlying position $p\in[0,1]$ in the distribution of 
Monte Carlo statistics is close to $0$ or $1$.
If we are 95\% confident that $p$ is within 0.025 of $0$ or $1$, then we claim the probability of 
the WMAP statistic happening by chance is sufficiently small to reject the hypothesis.

We find the WMAP data to have maxima that are significantly colder and minima that are significantly warmer than
predicted by Monte Carlo simulation.  For almost all simulations, we have 95\% confidence 
that the mean of the WMAP hot spots or cold spots is in a 5\% tail of the Monte Carlo distribution.
In one case, we are 99\% confident the maxima statistic is in a 1\% tail.
Since we find the same lack of extreme temperature when we use the directly measured WMAP power spectrum, we
are not simply restating that the WMAP power spectrum has a lack of power at large angular scales.
The effect is independent of the Galactic mask or power spectrum used.  

We also find some anisotropy between the ecliptic north and south hemispheres.
The WMAP data in northern hemisphere have a low variance statistic (95\% confident 
that the variance statistic is in a 5\% tail).   In one case, we are 99\% confident the
variance of the maxima is in a 1\% tail.
There is less asymmetry between the north and south
Galactic hemispheres.

Our results may not be a detection of primordial non-Gaussianity.
They could still be an effect of the WMAP instrument or data pipeline not modeled in our simulations
or an as yet undiscovered foreground.
Our result is still highly significant.
We have detected something, whether it is primordial non-Gaussianity or some other 
effect in the data.  
Having anomalous mean temperature values for
the maxima and minima in both the north and south ecliptic hemispheres 
is unlikely to occur 
if the WMAP data were consistent with theoretical expectations.
We will present a complete treatment of the one- and two-point extrema statistics for the WMAP data set
in a future publication.

\acknowledgments
Some of the results in this Letter have been derived using the HEALPix 
package 
(G{\'o}rski et al.\ 1999).
We would like to thank D. Spergel and O. Dor{\'e} for reading our manuscript.
This work was partially supported by the University of Illinois.


\end{document}